\documentclass[prd,10pt,amsmath,amssymb,aps,nofootinbib,showkeys,superscriptaddress,showpacs,notitlepage]{revtex4-1}
\usepackage{amssymb}
\usepackage[utf8]{inputenc}
\usepackage{amsmath}
\usepackage{amsfonts}
\usepackage{geometry}
\usepackage{color}
\usepackage{graphicx}
\usepackage{verbatim}

\def\sn{{\rm sn}}

\begin{document}

\title{Analytic non-homogeneous condensates in the $(2+1)$-dimensional
Yang-Mills-Higgs-Chern-Simons theory at finite density}

\author{Fabrizio \surname{Canfora}}
\email[]{canfora@cecs.cl}
\affiliation{Centro de Estudios Cient\'{\i}ficos (CECS), Casilla 1469,
Valdivia, Chile}

\author{Daniel \surname{Flores-Alfonso}}
\email[]{daniel.flores@cinvestav.mx}
\affiliation{Departamento de F\'{\i}sica, CINVESTAV-IPN, A.P. 14-740, C.P. 07000, Ciudad de M\'exico, Mexico}

\author{Marcela \surname{Lagos}}
\email[]{marcela.lagos@uach.cl}
\affiliation{Instituto de Ciencias F\'{\i}sicas y Matem\'{a}ticas,
Universidad Austral de Chile, Valdivia, Chile.}

\author{Aldo \surname{Vera}}
\email[]{aldo.vera@uach.cl}
\affiliation{Instituto de Ciencias F\'{\i}sicas y Matem\'{a}ticas,
Universidad Austral de Chile, Valdivia, Chile.}

\begin{abstract}
We construct the first analytic examples of non-homogeneous condensates in 
the Georgi-Glashow model at finite density in $(2+1)$ dimensions. The non-homogeneous condensates,
which live within a cylinder of finite spatial volume, possess a novel
topological charge that prevents them from decaying in the trivial vacuum. Also the 
non-Abelian magnetic flux can be computed explicitly. These solutions exist
for constant and non-constant Higgs profile and, depending on the length of the cylinder,
finite density transitions occur.  
In the case in which the Higgs profile is not constant, the full system of coupled field equations reduce to the Lam\'e equation 
for the gauge field (the Higgs field being an elliptic function).
For large values of this length, the energetically favored
configuration is the one with a constant Higgs profile, while, for small values, it is the one with non-constant Higgs profile.
The non-Abelian Chern-Simons term can also be included without spoiling the integrability properties of these
configurations. Finally, we study the stability of the solutions under a particular type of perturbations.
\end{abstract}

\maketitle

\newpage

\tableofcontents


\section{Introduction}


One of the most challenging open problems in theoretical and experimental
investigations in Quantum Chromodynamics (QCD) is to determine the phases
diagram at finite density and temperature, and especially, to shed light on
the confinement mechanism. Asymptotic freedom in the ultraviolet (UV) supports the melting
of Hadrons at high energies when Quarks and Gluons should be liberated, and
relativistic heavy-ion colliders allowed to realize high temperature
deconfined hadronic matter \cite{cabibbo}. This phase is relevant, for
instance, in the analysis of the core of compact stars (see for instance~%
\cite{2}, \cite{3}, \cite{4}, \cite{5} and \cite{6}).
Unfortunately both, in any heavy-ion experiment and in the core of
compact stars, QCD physics is dominated by the non-perturbative effects (see 
\cite{7} and references therein). In order to get insight on these
difficult problems lattice QCD (LQCD) simulations are effective \cite{15}, 
\cite{16}, \cite{19}, \cite{20}, \cite{21}, \cite{22}, \cite{23}%
, \cite{24}.

A quite remarkable discovery in this area is the appearance of
non-homogeneous condensates at finite density in the QCD phase diagram \cite%
{80}, \cite{81}, \cite{82}, \cite{83}, \cite{84}, \cite{85}, 
\cite{86}, \cite{87}, \cite{88} (see also \cite{89}, \cite{90}, 
\cite{91}, \cite{92}, \cite{93}, \cite{94}, \cite{95}, \cite{96}%
, \cite{98}, \cite{99}, \cite{100}, \cite{101}, \cite%
{102} and \cite{103}). This bold statement, also supported by the strong
phenomenological evidences favoring the so-called \textit{pasta phase}, has
been verified analytically in effective models in $(1+1)$ dimensions such as
the Gross-Neveu model \cite{pasta1}. A non-homogeneous pionic phase, usually
called Chiral Soliton Lattice (CSL), supported by strong external fields
is also possible. On the other hand, in the CSL the fact that the order
parameter only depends on one space-like coordinate prevents the CSL itself
from having a non-trivial topological charge and therefore the presence of a
strong external field is needed for stability reasons. In fact, recently it
has been shown that in the Skyrme model, which represents the low energy
limit of QCD in the 't Hooft expansion (see \cite{skyrme}, \cite{Witten}, 
\cite{witten0}, \cite{ANW}, \cite{bala0}, \cite{Bala1}), possesses analytic
crystal-like solutions with non-vanishing topological charge \cite{gaugsk}, 
\cite{LastUS1} (see also \cite{56b}, \cite{56}, \cite{LastUS2}, \cite%
{Cacciatori:2021neu} and \cite{LastUS3}). These configurations can be
interpreted as a topologically non-trivial CSL which are very much like a 
\textit{topologically protected version of the Larkin–Ovchinnikov–Fulde–Ferrell (LOFF) states} appearing in
normal superconductors \cite{LOFF1yLOFF2}. Moreover, perhaps surprisingly,
the subleading corrections to the Skyrme model in the 't Hooft expansion
(see \cite{marleau1}, \cite{Gudnason:2017opo}) do not spoil such analytic
solutions which ``resist" almost unchanged at any order in the large \textbf{%
N}$_{c}$ expansions \cite{LastUS1}.

Thus, as there are so many evidences of non-homogeneous and topologically
non-trivial condensates at finite density appearing in the low energy limit
of QCD at any order in the 't Hooft expansion, the natural question is: 
\textit{can we find this kind of topologically non-homogeneous condensates
directly in Yang-Mills theory?} This is a really fundamental question, as all
the relevant non-perturbative configurations of Yang-Mills theory, which are
important to understand the confinement mechanism, have been constructed in
``infinite space" (see \cite{manton}, \cite{shifman1}, and 
\cite{shifman2} for detailed pedagogical reviews), while the behavior of
these non-perturbative configurations living at finite density is largely
unknown (despite its huge interest in many applications). In this respect,
one of the main issues (which is discussed in this paper) yet to be
properly understood is how topologically non-trivial configurations react to
non-trivial boundary conditions at finite volume. Here, we analyze what happens when a finite amount of non-Abelian topological
charge related to gluons (as well as to the Higgs field) is forced to live
within a finite volume.

The simplest non-trivial case on which we focus here is
Yang-Mills-Higgs theory (Georgi-Glashow model with $SU(2)$ gauge group) in $%
(2+1)$ dimensions: this is a non-trivial interacting and confining gauge
theory. Needless to say, the issue of color confinement in non-Abelian gauge
theories is one of the most important and difficult open problems in modern
particle physics; especially, but not only in $(3+1)$ dimensions. The
available theoretical tools have not solved the $(3+1)$-dimensional problem
but it is pretty clear that non-perturbative configurations such as
monopoles, instantons and non-Abelian vortices play a fundamental role (see 
\cite{greensite}\ for a detailed review). On the other hand, non-Abelian
gauge theories in $(2+1)$ dimensions are still confining and interacting
theories but are much better understood from the analytic viewpoint. That is
why the Yang-Mills-Higgs theory is worth to be further investigated. The
insightful qualitative picture provided by Feynman \cite{feynman81} (based
on his early works on superfluidity in \cite{feynmanHe}) together with the
pioneering results of Polyakov on the role of monopoles \cite{polyakov} shed
considerable light on the important role that non-perturbative configurations play
in the $(2+1)$-dimensional confinement mechanism (see also \cite{kovner1}, 
\cite{kovner2}\ and references therein). The Hamiltonian approach to
Yang-Mills theory in $(2+1)$ dimensions has also provided remarkable results
on the mass gap, the string tension and the Glueball spectrum (see \cite%
{nair1}, \cite{nair2}, \cite{nair3}, \cite{leigh1}\ and references therein),
and the agreement of the results with the lattice approach is very good (see 
\cite{lattice}, \cite{lattice2}, \cite{lattice3} and references therein).

However, in the $(2+1)$-dimensional Georgi-Glashow model there are very few
analytic results on non-perturbative configurations at finite density. Here
it is worth to remind that the Polyakov discovery in \cite{polyakov} (see
also \cite{kovner1} and \cite{kovner2}) is based on the well-known 't
Hooft-Polyakov monopoles in $(3+1)$-dimensional Yang-Mills-Higgs theory
interpreted as Euclidean solutions\footnote{%
In other words, the 't Hooft-Polyakov monopoles in $(3+1)$-dimensional
Yang-Mills-Higgs theory play the role of instantons in the $(2+1)$%
-dimensional Georgi-Glashow model in $\mathbb{R}^{3}$.} in the
three-dimensional Euclidean flat $\mathbb{R}^{3}$. However, basically no
other non-perturbative configurations have been constructed analytically in
the literature in $(2+1)$-dimensional non-Abelian gauge theories. Even less
is known about how genuine non-perturbative configurations of these $(2+1)$%
-dimensional models react to the presence of non-trivial boundary conditions
such as finite volume effects. It is usually assumed that the presence of
finite volume effects (and, more generically, of non-trivial boundaries)
makes the field equations of the Yang-Mills-Higgs theory (which are already
by themselves a very hard nut to crack) even more difficult to solve since,
for instance, the usual spherical hedgehog ansatz cannot be used.

However, a systematic method to construct a generalized hedgehog ansatz
which is not spherically symmetric but keep all the other nice properties of
the usual hedgehog ansatz alive has been developed in \cite{gaugsk}, \cite%
{LastUS1}, \cite{56b}, \cite{56}, \cite{LastUS2}, \cite{Cacciatori:2021neu}, \cite{LastUS3} for the
Skyrme model, and such strategy has been proven useful also in the
Einstein-Yang-Mills case in \cite{ourYM1}, \cite{ourYM2} and \cite{ourYM3}; and generalizations thereof~\cite{ourYM4}.
In the present case, we generalize this technique to the case in which
the Yang-Mills-Higgs model is analyzed within a flat region of finite
spatial volume. We construct the first genuine analytic examples of
non-homogeneous and topologically non-trivial condensates in the
Georgi-Glashow model and in the Yang-Mills-Higgs-Chern-Simons theory in $%
(2+1)$ dimensions. Such solutions possess a novel topological charge and have
non-vanishing non-Abelian magnetic flux. Moreover, many relevant physical properties can
be computed explicitly (such as the energy density, the total energy, the
pressure and so on) in terms of the volume, the coupling constants and the
topological charge.

The paper is organized as follows: In Section 2, we introduce the
Yang-Mills-Higgs-Chern-Simons theory together with the general
parameterization for the fundamental fields. In Section 3, we propose our
ansatz and we show that analytic non-homogeneous condensates can be constructed for the pure
Yang-Mills theory. In Section 4, we construct non-homogeneous condensates in
the Georgi-Glashow model. In Section 5, we study the stability of the
solutions. In Section 6, we extend our results to the
Yang-Mills-Higgs-Chern-Simons theory. Finally, in the last section we draw some conclusions.


\section{The model}


In this section, we briefly review the Yang-Mills-Higgs-Chern-Simons theory
and display the general parameterization for the fundamental fields that
allows the construction of analytic non-homogeneous condensates.

\subsection{Yang-Mills-Higgs-Chern-Simons theory}

The Yang-Mills-Higgs-Chern-Simons theory in $(2+1)$ dimensions is defined by
the action 
\begin{equation}
I=\int d^{3}x\sqrt{-g}\biggl(\frac{1}{2e^{2}}\text{Tr}(F_{\mu \nu }F^{\mu
\nu})+\eta \left[ \frac{1}{4} \text{Tr}(D_{\mu }\varphi D^{\mu }\varphi)
-V\left( \varphi \right) \right] \biggl)\ + \mu \int d^{3}x\text{Tr}\left(
AdA+\frac{2}{3}A^{3}\right) \ ,  \label{I}
\end{equation}
where 
\begin{gather}
F_{\mu \nu }=\partial _{\mu }A_{\nu }-\partial _{\nu }A_{\mu }+[A_{\mu
},A_{\nu }]\ , \quad A=A_{\mu }dx^{\mu }=A_{\mu }^{j}t_{j}dx^{\mu }\ , 
\notag \\
V\left( \varphi \right) =\frac{\gamma }{4}\left(vev^2-|\varphi| ^{2}\right)
^{2}\ , \quad|\varphi|^2=-\frac{1}{2}\text{Tr}(\varphi ^{2})\ , \quad
t_{j}=i\sigma _{j}\ .  \label{V}
\end{gather}
Here $e$ and $\gamma$ are the coupling constants of Yang-Mills theory and of the Higgs potential, respectively, while, $vev$ is the
corresponding vacuum expectation value and $\mu$ is the Chern-Simons
coupling. In the previous equation, $\eta$ is either $0$ or $1$ depending
on whether one is only interested in pure Yang-Mills theory or in the
Georgi-Glashow model. The matrices $t_j$ are the generators of the $SU(2)$
group and $\sigma _{j}$ are the Pauli matrices. Lastly, we mention that $%
\varphi$ is the Higgs field in the adjoint representation and the covariant
derivative acts as 
\begin{equation*}
D_{\mu }\varphi =\partial _{\mu }\varphi +[A_{\mu },\varphi ]\ .
\end{equation*}
Varying the action w.r.t the fields $A_\mu$ and $\varphi$ we obtain the
field equations of the Yang-Mills-Higgs-Chern-Simons theory: 
\begin{gather}  \label{Eq2}
\nabla _{\nu }F^{\mu \nu }+[A_{\nu },F^{\mu \nu }]+\frac{\eta e^{2} }{4}%
[\varphi ,D^{\mu }\varphi ]\ + \frac{1}{2}\mu
e^2\varepsilon^{\alpha\beta\mu}F_{\alpha \beta }=\ 0\ , \\
D_{\mu }D^{\mu }\varphi + \gamma(vev^2-|\varphi|^2)\varphi=\ 0\ . \label{EqH}
\end{gather}
On the other hand, the energy-momentum tensor is given by 
\begin{equation*}
T_{\mu\nu}=-\frac{2}{e^{2}}\text{Tr}\biggl(F_{\mu \alpha }F_{\nu}{}^\alpha -%
\frac{1}{4}g_{\mu \nu }F_{\alpha \beta }F^{\alpha \beta }\biggl)-\frac{\eta}{2}  %
\text{Tr}\biggl(D_{\mu }\varphi D_{\nu }\varphi -\frac{1}{2}g_{\mu \nu
}D_{\alpha }\varphi D^{\alpha }\varphi \biggl) - g_{\mu\nu} V(\varphi) \ .
\end{equation*}%
For non-Abelian configurations, there are at least two possible definitions of magnetic flux. One option can be found in \cite{Abbott} and is based on the 
asymptotic symmetries of the field configurations and the existence of a normalized and covariantly constant Isospin vector. However, this approach does not apply to our configurations since $n^a$ in Eq. \eqref{generic2} (which is the natural choice of normalized unit vector in the internal space) is not covariantly constant. Hence, we use the following standard definition for the non-Abelian magnetic flux:
\begin{equation} 
\Psi ^{a}_{\text{M}} =\int d^{2}x\varepsilon ^{ij}F_{ij}{}^{a} \ .  \label{magneticflux}
\end{equation}
It is worth emphasizing that the importance of the Chern-Simons term in
combination with the Yang-Mills action has already been  disclosed in the
pioneering papers \cite{deser1}, \cite{deser2}, and such a combination is
also relevant in the analysis of QCD at high temperatures. The reason is
that in that regime QCD can be described as an effective three-dimensional
gauge theory in which (after integrating out the Fermions) the Chern-Simons
term shows up (see, for instance, \cite{27s}, \cite{28s}, and a detailed
review in \cite{3s}). To the best of authors' knowledge, there is no analytic
solution with non-trivial topological properties in the
Yang-Mills-Higgs-Chern-Simons theory.

\subsection{General parameterization}

One of the main motivations of the present analysis is to understand whether
or not interacting non-Abelian gauge theories possess non-homogeneous and
topologically non-trivial condensates at finite density, as it happens in
many of the low energy descriptions of QCD. The most natural way to take
into account finite volume effects is to use the metric defined below: 
\begin{equation}
ds^{2}=-dt^{2}+R^{2}dr^{2}+L^{2}d\phi ^{2}\ ,  \label{metric1}
\end{equation}%
where $R$ and $L$ are positive constants with dimension of length, representing the
size of the cylinder in which we are analyzing the system\footnote{We would like to point out that the analysis of solitons in cylindrical geometries, like the one we use in our paper, is very common in the apporach of adiabatic continuity and resurgence theory see, e.g.,~\cite{cherman}. In this area, it is very useful that the volume of the space in which the solitons live is a free parameter which can be varied.}. Both $r$ and $%
\phi $ are dimensionless coordinates with the following ranges 
\begin{equation}
0 \leq \phi < 2\pi \ , \qquad r_i \leq r\leq r_f \ ,
\label{ranges}
\end{equation}%
and the cylinder volume becomes $V_c=2 \pi (r_f-r_i) L R$.
The following general parameterization for the Yang-Mills and Higgs
fields is a natural generalization of the successful ansatz developed to
analyze non-homogeneous condensates for the Skyrme model in \cite{gaugsk}, 
\cite{LastUS1}, \cite{LastUS2} and \cite{Cacciatori:2021neu}. Given $U(x)\in SU(2)$, where
\begin{gather}
U^{\pm 1}(x^{\mu })=\cos \left( \alpha \right) \mathbf{1}_{2}\pm \sin \left(
\alpha \right) n^{i}t_{i}\ ,\ \ n^{i}n_{i}=1\ ,  \label{generic1} \\
n^{1}=\sin \Theta \cos \Phi \ ,\ \ n^{2}=\sin \Theta \sin \Phi \ ,\ \
n^{3}=\cos \Theta \ ,  \label{generic2} \\
\alpha =\alpha (x^{\mu })\ ,\quad \Theta =\Theta (x^{\mu })\ ,\quad \Phi
=\Phi (x^{\mu })\ ,  \label{generic3}
\end{gather}%
the ans\"{a}tze for the fields $A_{\mu }$ and $\varphi $ read 
\begin{gather}
A_{\mu }=\lambda \left( x^{\mu }\right) U^{-1}\partial _{\mu }U\ ,
\label{generic5} \\
\varphi =h\left( x^{\mu }\right) n^{j}t_{j}\ \ .  \label{generic4}
\end{gather}%
It is worth to point out that meronic gauge fields appear as a particular
case of the above configurations when $\lambda =1/2$. As it is well-known
(see \cite{actor}\ and references therein), a very interesting feature of
meron-type configurations is that such configurations can only appear in
non-Abelian gauge theories. The reason is that, in Abelian gauge theories, a
gauge potential which is proportional to a pure gauge is itself a pure gauge%
\footnote{%
When $\lambda $ is constant and $A_{\mu }$ is an Abelian pure gauge
configuration ($A_{\mu }=\partial _{\mu }\vartheta$ where $\vartheta$ is a
gauge parameter) then we have $A_{\mu }=\lambda \partial _{\mu }\vartheta
\Rightarrow A_{\mu }=\partial _{\mu }\left( \lambda \vartheta \right) $. Thus,
in the Abelian case, meron-type configurations are trivial.} and therefore
is trivial. On the other hand, in non-Abelian gauge theories, it is possible
to construct gauge potentials which are proportional to pure gauge but which
are not pure gauge themselves; these are the merons. Thus, in a sense,
merons are genuine features of non-Abelian gauge theories.


\section{Analytic non-homogeneous gluonic condensates}


Here, we discuss how to construct the ansatz in pure
Yang-Mills theory in $(2+1)$ dimensions in order to describe non-homogeneous gluonic condensates.

\subsection{The ansatz}

In this section, we consider a flat space-time described by the metric in
Eq. \eqref{metric1} with a finite length in the $r$ direction. Moreover,
we consider the pure Yang-Mills case, taking $\eta=\mu = 0 $ in Eq. \eqref{I}%
. Let us begin by discussing the idea behind the construction of the gauge
field. Arguably, the most convenient ansatz for the non-Abelian gauge
potential in Eq. \eqref{generic5} is the following: 
\begin{equation}  \label{ex1}
A_{\mu }=\lambda \left( r \right) U^{-1}\partial _{\mu }U \ .
\end{equation}%
This widely
used choice is convenient because when $\lambda$ is either $0$ or $1$ the
gauge field is trivial as it either vanishes or becomes a pure gauge,
respectively. Thus, $\lambda $ carries (part of) the
responsibility to make the gauge field ``non-trivial", and the ``pure gauge
part" $U^{-1}\partial _{\mu }U$ plays an important role in determining the
non-Abelian fluxes.

\subsubsection{Example: The non-Abelian monopole}

For instance, in the usual $(3+1)$-dimensional case, the spherical hedgehog
ansatz is given by Eqs. \eqref{generic1}, \eqref{generic2}, \eqref{generic3} and \eqref{ex1}, with the following form for the $U$
field 
\begin{equation}
\alpha = \frac{\pi}{2} \ ,\quad \Theta =\theta \ ,\quad \Phi =\varphi \ .
\label{ex5}
\end{equation}%
Considering the metric for the space-time as\footnote{Notice that in this metric $r$ is a radial coordinate and is not to be confused 
with the $r$ coordinate we use throughout this paper, cf. Eq. \eqref{metric1}.} 
\begin{equation}
ds^{2} = -dt^{2}+dr^{2}+r^{2}\left( d\theta ^{2}+\sin ^{2}\theta d\varphi
^{2}\right) \ ,  \label{ex2}
\end{equation}
the Yang-Mills equations reduce to just one ODE for the profile (see \cite{manton} for details). In this case, when the function $%
\alpha $ defined in Eq. (\ref{ex5}) is constant the magnetic flux is determined by the two-form $\Omega =d\Theta
\wedge d\Phi$, where $\Theta$ and $\Phi$ are the two functions appearing in
the ``Isospin vector" $n^{i}$\ in Eq. (\ref{generic2}). In other words, the
magnetic flux is non-vanishing across the two-dimensional surfaces
determined by the condition 
\begin{equation}
\Omega \neq 0 \ .  \label{cond1}
\end{equation}
Notice that with the choice in Eqs. \eqref{generic1}, \eqref{generic2}, \eqref{generic3} and \eqref{ex5}
one gets the usual magnetic flux of a spherical magnetic monopole. Moreover,
very similar arguments also hold in the case of electric fluxes. On the
other hand, the choice $\alpha =\text{const}$, is not mandatory. In
particular, one could consider an ansatz where $\Theta$ is constant and $%
\alpha$ is not. In this situation, non-trivial fluxes require $\Omega
^{\prime }=d\alpha \wedge d\Phi \ \neq 0$.

Indeed, in the following sections, we show that one can easily construct two
equivalent ans\"atze; one with $\alpha =\text{const}$ and $\Omega \neq 0$,
while, the other has $\Theta =\text{const}$ and $\Omega^{\prime }\neq 0$. The
choosing between these ans\"atze, in the case of pure Yang-Mills theory, is arbitrary. However, in the case of the Georgi-Glashow model, the
field equations are simpler with the choice $\Theta =\text{const}$, as we present further below.

\subsubsection{The first ansatz, $\protect\alpha=\text{const}$}

The most obvious ansatz in the family defined in Eqs. (\ref{generic1}), (\ref%
{generic2}), (\ref{generic3}) and (\ref{generic5}) corresponds to 
\begin{equation}  \label{opt1.1}
\alpha (x^{\mu }) = \frac{\pi }{2}\ ,\quad \Theta (x^{\mu })=\Theta (r)\
,\quad \Phi (x^{\mu }) = p\left( \frac{t}{L}-\phi \right) \ , \qquad p \in \mathbb{Z} \ ,
\end{equation}%
which satisfy the condition in Eq. (\ref{cond1}) to have a non-vanishing
magnetic flux. Here $p$ must be an integer in order to satisfy the periodicity condition in the $\phi$ direction of the field strength and the energy-momentum tensor (see Appendix). This is a non-spherical generalization of the usual hedgehog ansatz.
Note that this ansatz contains a light-like function $\Phi$ that allows to considerably reduce the field equations, as seen below, and it was one of the key ingredients to construct non-homogeneous condensates in the Skyrme model \cite{gaugsk}, \cite{LastUS1}.   

The $(2+1)$-dimensional Yang-Mills field equations corresponding to
the above choice reduces to the following two coupled non-linear ODEs for the
functions $\Theta (r)$ and $\lambda (r)$: 
\begin{equation}
\Theta ^{\prime \prime }+ \cot(\Theta) \Theta ^{\prime 2}+\frac{3}{2}\frac{%
(2\lambda -1)}{\lambda (\lambda -1)}\lambda ^{\prime }\Theta ^{\prime }=0\ ,
\label{opt1.3}
\end{equation}
\begin{equation}
\lambda ^{\prime \prime }+2\lambda (\lambda -1) \tan(\Theta) \Theta ^{\prime
\prime }+2\left(2 \cot(2 \Theta)-\csc{(2\Theta)}+3\tan(\Theta)\lambda
\right) \lambda ^{\prime }\Theta ^{\prime }-4\lambda \left( \lambda
-1\right) \Theta ^{\prime 2}=0 \ .  \label{opt1.4}
\end{equation}

\subsubsection{The second ansatz, $\Theta=\text{const}$}

The second choice in the family defined in Eqs. (\ref{generic1}), (\ref%
{generic2}), (\ref{generic3}) and (\ref{generic5}) corresponds to taking 
\begin{equation}  \label{opt2.1}
\alpha (x^{\mu }) = \alpha (r)\ ,\quad \Theta (x^{\mu })=\frac{\pi }{2}\ ,
\quad \Phi (x^{\mu }) = p\left( \frac{t}{L}-\phi \right) \ , \qquad p \in \mathbb{Z} \ ,
\end{equation}%
which satisfies $\Omega^{\prime }\neq0$; the condition equivalent to Eq. (\ref%
{cond1}) for the magnetic flux to be non-vanishing. In this case, the $(2+1)$%
-dimensional Yang-Mills field equations reduce to the following two coupled
non-linear ODEs for $\alpha (r)$ and $\lambda (r)$: 
\begin{equation}
\alpha ^{\prime \prime }+ \cot(\alpha) \alpha ^{\prime 2}+\frac{3}{2}\frac{%
(2\lambda -1)}{\lambda (\lambda -1)}\lambda ^{\prime }\alpha ^{\prime }=0\ ,
\label{opt2.3}
\end{equation}%
\begin{equation}
\lambda ^{\prime \prime }+2\lambda (\lambda -1) \tan(\alpha) \alpha ^{\prime
\prime }+2\left(2 \cot(2 \alpha)-\csc{(2\alpha)}+3\tan(\alpha)\lambda
\right) \lambda ^{\prime }\alpha ^{\prime }-4\lambda \left( \lambda
-1\right) \alpha ^{\prime 2}=0 \ .  \label{opt2.4}
\end{equation}%
Evidently, the two options are equivalent, in a sense, as one can see from
the comparison of the field equations in Eqs. \eqref{opt1.3}, \eqref{opt1.4} and Eqs. 
\eqref{opt2.3}, \eqref{opt2.4}. However, in the case of the Georgi-Glashow theory
(which we analyze in the next sections) the ansatz in Eq. \eqref{opt2.1}
leads to simpler field equations. Henceforth, we consider Eq. \eqref{opt2.1}
within the family of configurations defined by Eqs. \eqref{metric1}, \eqref{ranges}, \eqref{generic1},
\eqref{generic2}, \eqref{generic3} and \eqref{generic5}.

\subsection{Gluonic condensates}

At a first glance, the task to find analytic solutions of the field
equations in Eqs. (\ref{opt2.3}) and (\ref{opt2.4}) seems to be completely
hopeless because not only the field equations are non-linear (as one would
expect in the Yang-Mills theory), but they are also coupled and there is no
obvious BPS trick in this case. However, this is not the case. We now show
that it is indeed possible to find exact solutions of that system of non-linear ODEs. The best strategy to construct analytic non-homogeneous
gluonic condensates is, first of all, to think that the function $\lambda $
depends on $\alpha $%
\begin{equation}
\lambda =\lambda (\alpha )\ ,  \label{strat1}
\end{equation}%
so that $\lambda $ depends on the coordinate $r$ only through $\alpha $.

Secondly, we have to ask the following question: \textit{How should }$%
\lambda $\textit{\ depend on }$\alpha $\textit{\ in such a way that after
replacing such }$\lambda =\lambda (\alpha )$\textit{\ in the two field
equations then Eqs. (\ref{opt2.3}) and (\ref{opt2.4}) reduce to just one
equation for }$\alpha (r)$?

Although, a priori, it is not obvious at all that such a functional
dependence of $\lambda $ on $\alpha $ with the above property really exists,
it is a direct computation to show that the expression here below does the
job 
\begin{equation}
\lambda (\alpha )=\frac{1}{2}\left( 1\pm \frac{\cos \alpha }{\sqrt{\cos
^{2}\alpha +k}}\right) \ ,  \label{changeofvariable1}
\end{equation}%
where the auxiliary parameter $k$, which is a useful by-product of our
analysis, is an integration constant. Indeed, a direct computation reveals
that if $\lambda $ depends on $\alpha $ as in Eq. (\ref{changeofvariable1})
the Yang-Mills field equations reduce to the following ODE for $\alpha (r)$: 
\begin{equation}
\alpha ^{\prime \prime }+\left[ \cot \alpha +\frac{3\sin \alpha \cos \alpha 
}{\cos ^{2}\alpha +k}\right] \alpha ^{\prime 2}=0\ .  \label{gluonic1}
\end{equation}%
It is noteworthy that $k=0$ in Eq. (\ref{changeofvariable1}) yields the
trivial ``pure gauge solutions", namely $\lambda =1,0$, as it is clear from
Eq. (\ref{generic5}). On the other hand, the limit $k\rightarrow \infty $
yields $\lambda =\frac{1}{2}$, thus meronic configurations can be obtained
in this limit. The plus and then the minus sign branch of Eq. (\ref%
{changeofvariable1}) are plotted below. 
\begin{figure}[ht]
\centering
\includegraphics[scale=0.5]{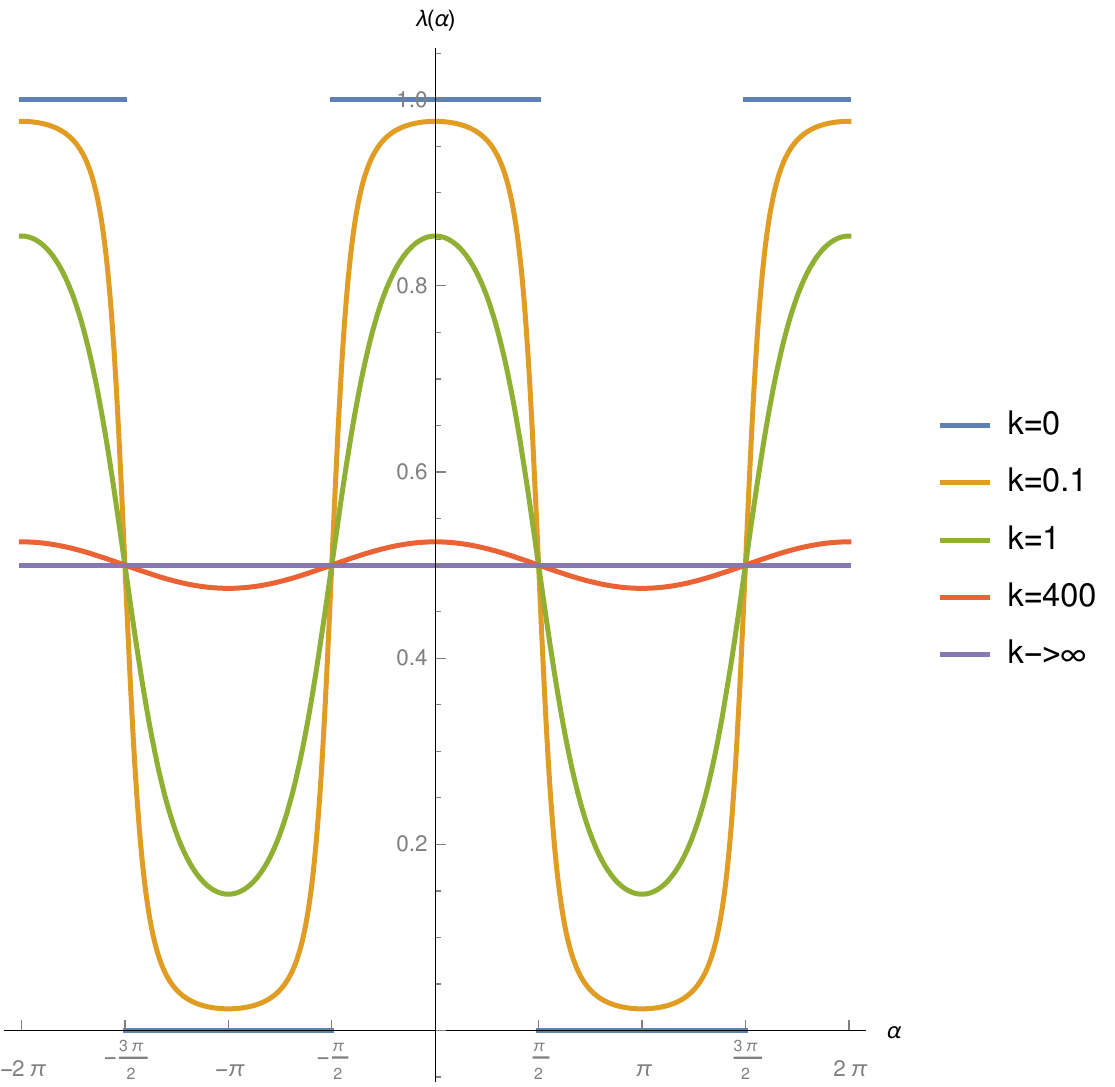} \ \qquad %
\includegraphics[scale=0.5]{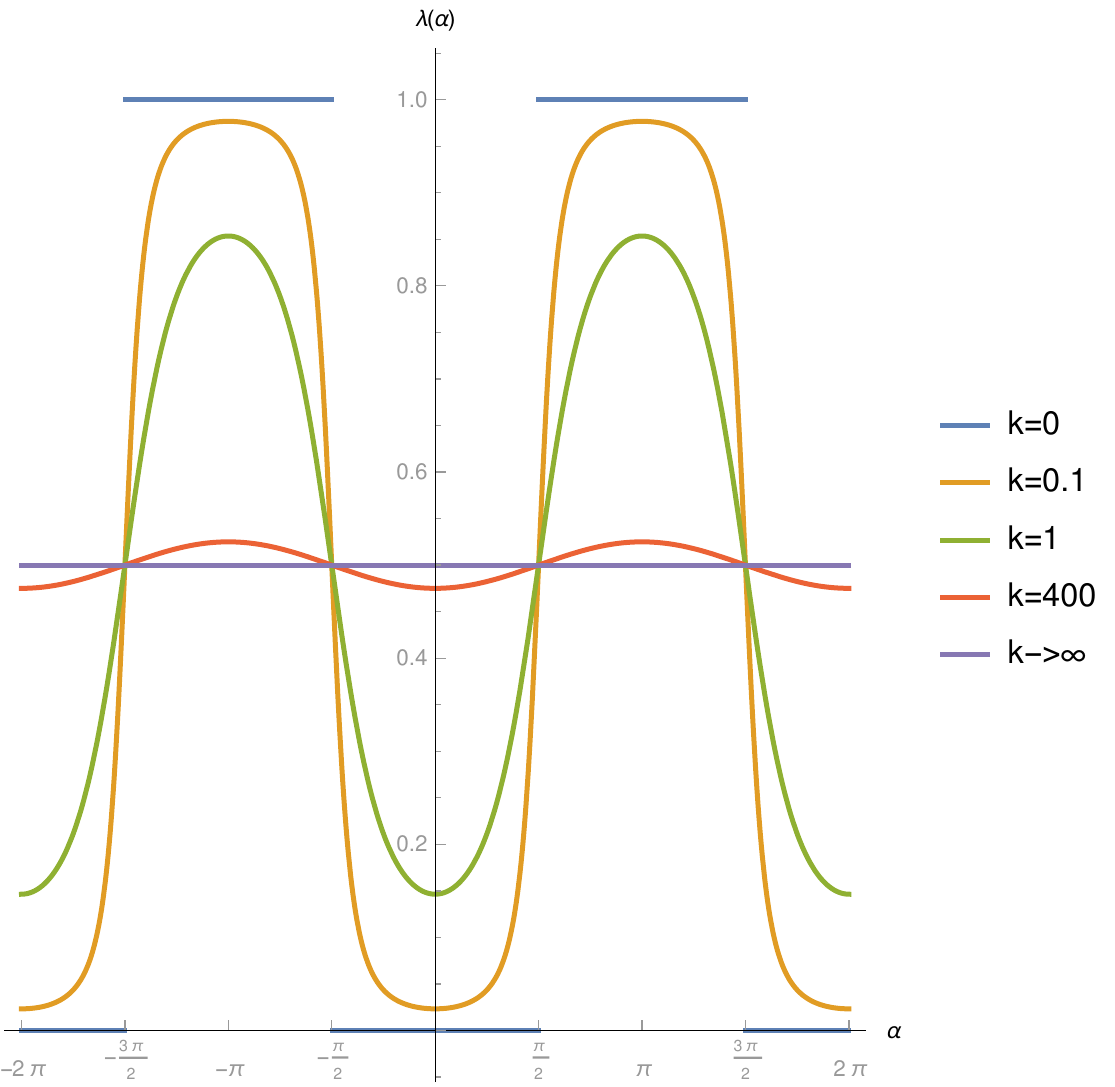}
\caption{The positive and negative branches of $\lambda(
\alpha)$ in Eq. \eqref{changeofvariable1} are plotted for different values of $k$.}
\label{branches}
\end{figure}
Then, with the change of variable in Eq. \eqref{changeofvariable1},
our task has been reduced to solving a single ODE for $\alpha (r)$ in Eq. (%
\ref{gluonic1}). Interestingly enough, such an ODE is solvable since it can
be reduced to the following quadrature 
\begin{gather}  \label{gluonic1.1}
\left[ \frac{\sin ^{2}(\alpha) }{2(\cos^{2}(\alpha) +k)^{3}}\right] \alpha
^{\prime 2} = E_{0} \ \ , \\
\frac{d\alpha}{\rho \left( \alpha ,E_{0}\right)} = \pm dr\ , \qquad \rho
\left( \alpha ,E_{0}\right) = \frac{\left( 2E_{0}\right) ^{1/2}(\cos
^{2}(\alpha) +k)^{3/2}}{\sin(\alpha) } \ ,  \label{gluonic2}
\end{gather}
where $E_{0}$ is an integration constant. Moreover, Eq. \eqref{gluonic1}
(or, equivalently, Eq. \eqref{gluonic2}) can be solved explicitly, however,
in order to compute all the relevant quantities such as the total energy,
pressure and so on, the expression in Eq. \eqref{gluonic2} is sufficient.

Summarizing, the function $\alpha$, determined explicitly by the quadrature
in Eqs. \eqref{gluonic1.1} and \eqref{gluonic2}, in turn determines the
dependence of $\lambda $ on $r$; through Eq. \eqref{changeofvariable1}.
These two functions are the analytic solutions of the $(2+1)$-dimensional
Yang-Mills equations in Eqs. \eqref{opt2.3} and \eqref{opt2.4} on the
cylinder defined in Eq. \eqref{metric1}.

\subsection{The energy of the condensate}

The energy density $T_{00}$ (which in the case of the ansatz in Eq. (\ref%
{opt2.1}) within the family of configurations defined in Eqs. (\ref{generic1}%
), (\ref{generic2}), (\ref{generic3}) and (\ref{generic5})) reads 
\begin{equation}
T_{00}=\frac{4p^{2}\left[ 4\lambda ^{2}(\lambda -1)^{2}\alpha ^{\prime
2}+\lambda ^{\prime 2}\right] \sin ^{2}(\alpha )}{e^{2}R^{2}L^{2}}\ .
\label{gluonicenden1}
\end{equation}%
When one takes into account that $\lambda $ depends on $\alpha $ as in Eq. (%
\ref{changeofvariable1}) the energy density becomes 
\begin{gather}
T_{00}=\frac{\xi }{\rho \left( \alpha ,1\right) ^{2}}\alpha ^{\prime 2}\ ,
\label{gluonicenden2} \\
\rho \left( \alpha ,1\right) =\rho \left( \alpha ,E_{0}=1\right) \ ,\quad
\xi =\frac{2\left( kp\right) ^{2}(k+1)}{\left( LRe\right) ^{2}}\ ,  \notag
\end{gather}%
where $\rho \left( \alpha ,E_{0}\right) $\ has been defined in Eq. %
\eqref{gluonic2}. In this way, the total energy $E_{\text{tot}}$ of the
system can be written as%
\begin{equation}
E_{\text{tot}}=RL\int drd\phi T_{00} =2 V_c \frac{E_0 k^2 (k+1) p^2}{e^2 L^2 R^2}
    \ ,  \label{totalE}
\end{equation}
where $ V_c $ is the cylinder volume defined in Eq. \eqref{ranges}. It is worth to emphasize that these inhomogeneous gluonic condensates are non-perturbative, i.e.,  Eq. \eqref{totalE} is singular around $e=0$. Moreover, notice that the energy density turns out to be constant, and it
depends on the parameter $k$ in Eq. \eqref{changeofvariable1} as well as on the integration constant $E_0$ in Eq. \eqref{gluonic1.1}.
$E_0$ is determined through the relation
\begin{equation*}
r_f-r_i=\int_{\alpha(r_i)}^{\alpha(r_f)} \frac{1}{\rho(\alpha,E_0)}d\alpha \ , 
\end{equation*}
once the boundary conditions are chosen, while the allowed values of $k$ can be calculated by requiring that the non-Abelian magnetic flux be quantized. 
We compute the non-Abelian magnetic flux in the next section for the most interesting case of the Georgi-Glashow model, this for two reasons that we detail here. The first reason is that the inclusion of the Higgs field allows to derive a BPS bound, from which a topological charge naturally emerges, 
and the non-triviality of this quantity determines the appropriate boundary conditions for the $\alpha$ profile. 
The second reason is that the non-Abelian magnetic flux, both for the case with Higgs field and without the Higgs, are actually the same. 
We detail more about these points in the next section. 


\section{Analytic non-homogeneous condensates in the Georgi-Glashow model}


In this section, we construct non-homogeneous condensates in the
Georgi-Glashow model. Thus, we set $\eta =1$ and $\mu=0$, in Eq. \eqref{I}.
As it has been already emphasized, in this model our ansatz is the one in
Eq. \eqref{opt2.1} within the family of configurations defined in Eqs. \eqref%
{generic1}, \eqref{generic2}, \eqref{generic3}, \eqref{generic5} and \eqref{generic4}, and in the particular case when the Higgs field is
\begin{equation} \label{h}
 h=h(r) \ . 
\end{equation}

\subsection{Solving the field equations}

The $(2+1)$-dimensional Georgi-Glashow field equations for the ansatz defined
in Eqs. \eqref{metric1}, \eqref{ranges}, \eqref{generic1}, \eqref{generic2}, \eqref{generic3}, \eqref{generic5}, \eqref{generic4} and \eqref{h} reduce to the following three coupled non-linear ODEs for $\alpha (r)$, $\lambda (r)$ and $h(r)$:
\begin{equation}
\alpha ^{\prime \prime }+\cot (\alpha )\alpha ^{\prime 2}+\frac{3}{2}\frac{%
(2\lambda -1)}{\lambda (\lambda -1)}\lambda ^{\prime }\alpha ^{\prime }=%
\frac{e^{2}R^{2}}{4}\frac{\cot (\alpha) }{\lambda (\lambda -1)}h^{2}\ ,
\label{gg1}
\end{equation}%
\begin{equation}
\lambda ^{\prime \prime }+2\lambda (\lambda -1)\tan(\alpha) \alpha ^{\prime
\prime }+2\left[ 2\cot(2 \alpha)-\csc(2\alpha)+3\lambda\tan(\alpha) \right]
\lambda ^{\prime }\alpha ^{\prime }-4\lambda \left( \lambda -1\right)^2
\alpha^{\prime 2}=e^{2}R^{2}\lambda h^{2}\ ,  \label{gg2}
\end{equation}%
\begin{equation}
h^{\prime \prime }+ \gamma R^2 \left(vev^2 - h^{2}\right) h=0\ .  \label{gg3}
\end{equation}
We emphasize two remarkable features about the ansatz in Eq. (\ref{opt2.1}), which is a
generalization of the strategy developed in Refs.\footnote{%
In Refs. \cite{gaugsk}, \cite{LastUS1} and \cite{LastUS2}, where the Maxwell gauged Skyrme model was considered, the following issue
arose: is it possible to find an ansatz for the Skyrmion and the gauge field
in such a way that the ``gauge field disappears" from the Skyrme field
equations without the gauge field being trivial and at the same time keeping
the Baryon charge alive? The answer was proven to be affirmative.} \cite%
{gaugsk}, \cite{LastUS1}, \cite{LastUS2} for the Skyrme
model. Firstly, although we have a non-trivial Higgs field, a direct
computation shows that if $\lambda $ depends on $\alpha $, as in Eq. %
\eqref{changeofvariable1}, then Eqs. (\ref{gg1}) and (\ref{gg2}) reduce
again to just one ODE for $\alpha (r)$ 
\begin{equation}
\alpha ^{\prime \prime }+\left[ \cot \alpha +\frac{3\sin \alpha \cos \alpha 
}{\cos ^{2}\alpha +k}\right] \alpha ^{\prime 2}+\frac{e^{2}R^{2}}{k}\cot
\alpha (\cos ^{2}\alpha +k)h^{2}=0\ .  \label{gg2.1}
\end{equation}
Moreover, the energy density becomes 
\begin{equation}  \label{ED}
T_{00} = \frac{p^2 k^2(k+1)}{e^2R^2L^2} \frac{\sin^2(\alpha)}{%
(\cos^2(\alpha)+k)^3}\alpha^{\prime 2 }+ \frac{p^2(k+1)\cos^2(\alpha)}{%
L^2(\cos^2(\alpha)+k)}h^2 + \frac{1}{2R^2}h^{\prime 2 }+ \frac{\gamma}{4}%
(vev^2-h^2)^2 \ .
\end{equation}
Secondly, despite the facts that the ansatz is genuinely non-Abelian and
that the commutators between the Higgs field and the gauge field as well as
the commutators of the gauge field with itself are non-vanishing (see the
Appendix), all the terms in the field equations for the Higgs field which
could, in principle, couple the Higgs profile $h(r)$ with the gauge field
profiles $\lambda (r)$ and $\alpha (r)$ actually vanish. The main technical
reason behind this simplification is that the fields $\alpha $, $\Theta $
and $\Phi $ in Eq. (\ref{generic3}) have been chosen in such a way that
\begin{equation*}
\partial _{\mu }\alpha \partial ^{\mu }\Phi =\partial _{\mu }\Theta \partial
^{\mu }\Phi =\partial _{\mu }\Phi \partial ^{\mu }\Phi =\partial _{\mu
}h\partial ^{\mu }\Phi =0\ .
\end{equation*}%
This ``decoupling property" is the Yang-Mills-Higgs generalization of the
approach used in the gauged Skyrme model minimally coupled
with the Maxwell field in \cite%
{gaugsk}, \cite{LastUS1}, \cite{LastUS2}. Due to this
decoupling property, Eq. (\ref{gg3}) can be reduced to the following
quadrature 
\begin{equation}
\frac{\left( h^{\prime }\right) ^{2}}{2}-\frac{\gamma R^{2}}{4}\left(
h^{2}-vev^{2}\right) ^{2}=\frac{I_{0}}{2}\qquad \Rightarrow \quad \pm dr\ =%
\frac{dh}{\sqrt{I_{0}+\frac{\gamma R^{2}}{2}\left( h^{2}-vev^{2}\right) ^{2}}%
}\ ,  \label{gg3.2}
\end{equation}%
where $I_{0}$ is an integration constant. Consequently, one can solve
explicitly the Higgs field in Eq. (\ref{gg3}) (or equivalently in Eq. %
\eqref{gg3.2}) in terms of a Jacobi elliptic function, that is 
\begin{equation}
h(r)=K_{0}\mathrm{sn}(u(r)-u_{0},\kappa )\ ,  \label{hexplicit}
\end{equation}%
with 
\begin{equation}
K_{0}=\kappa \sqrt{\frac{2}{\kappa^2 +1}}vev\ ,\quad \text{and}\quad u(r)=%
\frac{K_{0}}{\kappa }\sqrt{\frac{\gamma }{2}}Rr\ .  \label{Ku}
\end{equation}%
In the previous equations, the integration constants $u_{0}$ and $\kappa $ correspond to the phase
and the elliptic modulus, respectively. Without any loss of generality, we
can fix $u_{0}=0$ because it just corresponds to a shift in the $r$
coordinate and can be absorbed into a coordinate's redefinition. Moreover, inserting Eqs. \eqref{hexplicit} and \eqref{Ku} into \eqref{gg3.2} yields
\begin{equation}
 I_0=-\frac{\gamma R^2(vev)^4(1-\kappa^2)^2}{2(1+\kappa^2)^2},
\end{equation}
which makes manifest that $\kappa=1$ corresponds to $I_0=0$.

Therefore, we have reduced the problem to solve the three coupled non-linear
ODEs in Eqs. (\ref{gg1}), (\ref{gg2}) and (\ref{gg3}) to solve only Eq. (\ref%
{gg2.1}), where $h$ is explicitly known. Notice that there are three types of
possible solutions for the Higgs field. The simplest non-trivial solution
corresponds to taking the profile $h=\pm vev$ in Eq. \eqref{gg3}. The second
option is to consider a non-constant solution of Eq. \eqref{gg3.2} with $%
I_{0}=0$ (or $\kappa=1$ in Eq. \eqref{hexplicit}), which corresponds to a kink-type solution. Otherwise, for $I_0\neq 0$ in Eq. \eqref{gg3.2} the
solution is periodic. We consider these three possibilities in separate sub-sections.

At this point it is important to emphasize that despite the dramatic
simplification of the field equations that we have previously shown, the
configurations constructed here are genuinely non-Abelian. Indeed, not only
many of the commutators between the gauge potential and the Higgs fields are
non-vanishing (see the Appendix), but also one can see that meron-type
configurations (which only appear in non-Abelian gauge theories) are a
particular case of the present family of configurations; cf. Eq. (\ref{generic5}) and the
comments below it.

\subsection{Constant Higgs profile}

In this subsection we consider $h=\pm vev$, for which Eq. (\ref{gg2.1})
becomes 
\begin{equation}
\alpha ^{\prime \prime }+\left[ \cot \alpha +\frac{3\sin \alpha \cos \alpha 
}{\cos ^{2}\alpha +k}\right] \alpha ^{\prime 2}+\frac{e^{2}R^{2}\left(
vev\right) ^{2}}{k}\cot \alpha (\cos ^{2}\alpha +k)=0\ ,  \label{ggvev1}
\end{equation}%
while, the Higgs equation in Eq. \eqref{gg3} is automatically satisfied.
Quite interestingly, the above equation can be reduced to a quadrature since
Eq. (\ref{ggvev1}) possesses the following first integral: 
\begin{equation}
\left[ \frac{\sin ^{2}\alpha }{2(\cos ^{2}\alpha +k)^{3}}\right] \alpha
^{\prime 2}+\frac{e^{2}R^{2}\left( vev\right) ^{2}}{2k(\cos ^{2}\alpha +k)}%
=E_{0}\ ,  \label{ggvev2}
\end{equation}%
where $E_{0}$ is an integration constant\footnote{Indeed, it is easy to see that the
derivative of Eq. (\ref{ggvev2}) is proportional to Eq. (\ref{ggvev1}).}.
Thus, the complete set of field equations of the Georgi-Glashow model can be
reduced to the following quadrature:%
\begin{equation}
\frac{d\alpha }{\rho \left( \alpha ,E_{0}\right) }=\pm dr\ , \qquad \rho
\left( \alpha ,E_{0}\right) = \frac{\cos^2{\alpha}+k}{\sin{\alpha}} \sqrt{
2(\cos^2{\alpha}+k)E_0 -\frac{e^2R^2 (vev)^2}{k}} \ .  \label{ggvev3}
\end{equation}%
We also mention that the above quadrature can be explicitly solved in terms
of generalized elliptic integrals (see \cite{Elliptics}), although, one can
compute analytically relevant physical quantities, such as the energy and
the pressure, just using Eq. (\ref{ggvev3}), as shown below.

\subsubsection{Energy density and BPS bound}

The energy density in Eq. \eqref{ED} with $h=\pm vev$ is reduced to 
\begin{equation}
T_{00}=\frac{p^{2}(k+1)}{L^{2}}\left[ \frac{k^{2}\sin ^{2}(\alpha) }{%
e^{2}R^{2}(\cos ^{2}(\alpha) +k)^{3}}\alpha ^{\prime 2}+\frac{\left(
vev\right) ^{2}\cos ^{2}(\alpha) }{\cos ^{2}(\alpha) +k}\right] \ .
\label{ggvevbps1}
\end{equation}%
A very relevant feature of the above expression for the energy density and,
in fact, of the full energy-momentum tensor, is that it has the right
periodicity in $\phi $. In particular, we remind the reader that, in the case of gauge
theories, one has to require that physical gauge-invariant observables (such
as the energy density) must be periodic functions of $\phi $. However, the
gauge potential itself (which is not gauge-invariant) need not be so.

In this case, it is possible to derive a non-trivial BPS bound rewriting $%
T_{00}$ as%
\begin{equation}
T_{00}=\frac{p^{2}(k+1)}{L^{2}}\left[ \left( \varsigma \left( \alpha \right)
\alpha ^{\prime }\pm W\left( \alpha \right) \right) ^{2}\mp 2\varsigma
\left( \alpha \right) W\left( \alpha \right) \alpha ^{\prime }\right] \ ,
\label{ggvevbps2}
\end{equation}%
\begin{equation}
\varsigma \left( \alpha \right) =\frac{k\sin (\alpha )}{eR(\cos ^{2}(\alpha
)+k)^{3/2}}\ ,\quad W\left( \alpha \right) =\frac{\left( vev\right) \cos
(\alpha )}{\left( \cos ^{2}(\alpha )+k\right) ^{1/2}}\ .  \label{ggvevbps3}
\end{equation}%
Consequently (since the term $2\varsigma \left( \alpha \right) W\left(
\alpha \right) \alpha ^{\prime }$\ in Eq. (\ref{ggvevbps2}) is a total
derivative), the following BPS bound on the total energy, defined in Eq. %
\eqref{totalE}, can be derived 
\begin{equation}
E_{\text{tot}}\geq \left\vert Q^{H}\right\vert \ ,  \label{ggvevbps4}
\end{equation}%
where the topological charge is given by 
\begin{align}
Q^{H}& = \pm \frac{p^{2}(k+1)}{L^{2}}\times 2\pi RL\int_{r_i}^{r_f}2\varsigma
\left( \alpha \right) W\left( \alpha \right) \frac{d\alpha }{dr}dr\   \notag
\\
& = \pm \left. \frac{2\pi p^{2}k(k+1)vev}{e\,L\,(\cos ^{2}\alpha +k)}\right\vert
_{\alpha (r_i)}^{\alpha (r_f)}\ .  \label{ggvevbps5}
\end{align}%
The above bound can be saturated if and only if the following first order
equation is satisfied:%
\begin{equation}
\varsigma \left( \alpha \right) \alpha ^{\prime }=\pm W\left( \alpha \right)
\ .  \label{ggvevbps6}
\end{equation}%
It is a very non-trivial result that in the presence of a Higgs field the
BPS condition, here above, implies that the second order field equation in Eq. \eqref{ggvev1} is
satisfied.

Notice that the first-order equations in Eqs. \eqref{ggvev2} and \eqref{ggvevbps6} are compatible for a particular value
of the integration constant $ E_0 $, namely
\begin{equation*}
 E_0 \ = \ \frac{e^2R^2 (vev)^2}{2k^2} \ . 
\end{equation*}
While the solutions of the equation that comes from the saturation of the BPS bound in Eq. \eqref{ggvevbps6} only correspond
to a set of all the allowed solutions of Eq. \eqref{ggvev1}, all solutions of Eq. \eqref{ggvev1} are also solution of Eq. \eqref{ggvev2}.
Therefore, though the existence of a BPS bound that allows to solve the Yang-Mills-Higgs system
analytically is something clearly non-trivial, throughout the paper we do not refer to the
solutions that can be obtained from Eq. \eqref{ggvevbps6}, but rather to those general solutions of Eq.
\eqref{ggvev1} (or equivalently Eq. \eqref{ggvev2}). 

\subsubsection{Boundary conditions and topological charge}

Evaluating the topological charge in Eq. \eqref{ggvevbps6} at
the top and the bottom of the cylinder in the range defined in Eq. %
\eqref{ranges} we see that, in order to have a non-vanishing topological charge, we must
demand that, $\cos ^{2}{\alpha (r_i)}\neq \cos ^{2}{\alpha (r_f)}$. Then, suitable boundary conditions for the $\alpha 
$ profile are 
\begin{equation}
\alpha (r_i)=\frac{\pi }{2}\ ,\qquad \alpha (r_f)=0\ .  \label{bc3}
\end{equation}%
In fact, with the above boundary conditions the topological charge becomes 
\begin{equation} \label{QH}
Q^{H}= \pm \frac{2\pi p^{2}(vev)}{eL}\ .
\end{equation}%
Note that the appropriate boundary conditions for the $\alpha$ profile can be read directly from Eq. \eqref{ggvev3}. Indeed,
as we are looking for regular solutions it is necessary that $\alpha'$
must not have singularities or change sign, and this implies that $\alpha$ can only be extended in a length range from $0$ to $\frac{\pi}{2}$. 
Now, the integration constant $E_{0}$ is fixed through the relation 
\begin{equation*}
r_f-r_i=\pm \int_{0}^{\frac{\pi }{2}}\frac{1}{\rho (\alpha ,E_{0})}d\alpha \ ,
\end{equation*}%
with $\rho (\alpha ,E_{0})$ defined in Eq. \eqref{ggvev3}.

To the best of the authors' knowledge, the topological charge in Eqs. (\ref{ggvevbps4}%
), (\ref{ggvevbps5}) and \eqref{QH} is novel, non-trivial and useful. First of all, it
is novel since $Q^{H}$ does not coincide with the non-Abelian magnetic flux or the enclosed electric charge which, usually, play the role of
topological charges in non-Abelian gauge theories. It is non-trivial since
we can construct solutions with non-vanishing $Q^{H}$. It is useful since
the requirement to saturate the BPS bound in Eq. (\ref{ggvevbps4}) gives
rise to a first order condition, which implies the second order field
equations. These results are likely to be genuine finite density effects and
are, consequently, very relevant when analyzing the theory within a finite
volume.

\subsection{Non-constant Higgs profile: The general case}

Not surprisingly, the case in which the Higgs profile is non-constant is
considerably more difficult. Nevertheless, many analytic results can be
derived. The field equation for the profile $\alpha $ in this case is given
by 
\begin{equation}
\alpha ^{\prime \prime }+\left[ \cot \alpha +\frac{3\sin \alpha \cos \alpha 
}{\cos ^{2}\alpha +k}\right] \alpha ^{\prime 2}+\frac{e^{2}R^{2}}{k}\cot
\alpha (\cos ^{2}\alpha +k)h^{2}=0\ ,  \label{nonconstant}
\end{equation}%
where $h$ is a non-constant solution of Eq. (\ref{gg3.2}) defined in general
in Eq. \eqref{hexplicit}. In order to simplify the above equation it is
useful to introduce a new function of $\alpha $, which we denote by $%
\Gamma \left( \alpha \right) $, and is defined by the following relation: 
\begin{equation}
\frac{d\Gamma \left( \alpha \right) }{dr}=\mp \left[ \frac{\sin (\alpha )}{%
2(\cos ^{2}(\alpha )+k)^{3/2}}\right] \alpha ^{\prime }\qquad \Rightarrow
\quad \Gamma \left( \alpha \right) =\pm \frac{\cos (\alpha )}{2k\sqrt{\cos
^{2}(\alpha )+k}}\ ,  \label{newchange}
\end{equation}%
so that the profile $\alpha $ can be written\footnote{The idea of the change of variables between $\alpha $ and $\Gamma $ in Eqs. (%
\ref{newchange}) and (\ref{newchange2}) is the following. The first two
terms in Eq. (\ref{nonconstant}) are proportional to the second derivative
of the function $\Gamma (\alpha )$ in Eq. (\ref{newchange}). Hence, if one
uses Eq. (\ref{newchange2}) which expresses $\alpha $ in terms of $\Gamma $,
then the first derivative term in the field equation in Eq. (\ref%
{nonconstant}) disappears.} as 
\begin{equation}
\alpha =\arccos \left[ \biggl(\frac{k\Gamma ^{2}}{\frac{1}{4k^{2}}-\Gamma
^{2}}\biggl)^{\frac{1}{2}}\right] \ .  \label{newchange2}
\end{equation}%
Then one is left with the following simpler linear equation 
\begin{equation}
\Gamma ^{\prime \prime }-(e^{2}R^{2}h^{2})\Gamma =0\ .  \label{almostfinal}
\end{equation}%
Notice that by examining Eq. \eqref{newchange} and comparing it with Eq.
\eqref{changeofvariable1}, the above $\Gamma (\alpha )$ function is
proportional to our earlier defined $\lambda (\alpha )$ function 
\begin{equation}
\Gamma =\frac{2\lambda -1}{2k}\ .
\end{equation}%
From Eq. (\ref{newchange2}), the complete set of
field equations of the Georgi-Glashow model with a non-constant Higgs
profile has been reduced to just Eq. (\ref{almostfinal}) where $h$ is a
non-constant solution of Eq. (\ref{gg3.2}).

\subsubsection{Mapping with the Lam\'e equation}

We now move on to solving the only pending ODE to have a complete
solution of the Georgi-Glashow field equations.
Let us recall that the non-constant Higgs profile obeys Eq. %
\eqref{gg3.2}. In that equation, the integration constant $I_0$
characterizes the configurations period determined by $\kappa$, and the general
solution is an elliptic sine function as showed in Eq. \eqref{hexplicit}. In
this case, the system can be taken to solve a Lam\'e equation.

Notice that Eq. \eqref{hexplicit} is at its simplest in terms of the
variable $u$ defined in Eq. \eqref{Ku}. Thus, we transform %
Eq. \eqref{almostfinal} by considering 
\begin{equation}
y(u)=\Gamma(r(u)) \ ,
\end{equation}
so that it becomes 
\begin{equation}
\frac{\mathrm{d}^2y}{\mathrm{d} u^2}-l(l+1)\kappa^2\mathrm{sn}^2(u,\kappa)
\,y =0 \ ,  \label{lame}
\end{equation}
where $l$ has been defined so that it satisfies $l(l+1)=2e^2/\gamma$, for
which there are always solutions. This combination of the coupling
constants $e$ and $\gamma$ is always positive and so leads to real values of 
$l$. 

The ODE in Eq. \eqref{lame} is known as the Lam\'e equation and its solutions as
Lam\'e functions. Special situations arise when $l$ is an integer, however,
solutions always exist for general complex values of $l$, that would be acceptable to us. It is well-known
that the Lam\'e functions are a special case of Heun functions. We write the
general solution of Eq. \eqref{lame} as 
\begin{equation}
y(u)=c_1\,H\left(\frac{1}{\kappa^2},0,-\frac{l}{2},\frac{l+1}{2},\frac{1}{2},%
\frac{1}{2},\mathrm{sn}^2{u}\right)+c_2\,\mathrm{sn}{u}\,H\left(\frac{1}{%
\kappa^2},\frac{1+\kappa^2}{4\kappa^2},\frac{l+2}{2},\frac{1-l}{2},\frac{3}{2%
},\frac{1}{2},\mathrm{sn}^2{u}\right) \ ,  \label{soly}
\end{equation}
where every elliptic function has the same elliptic modulus $\kappa$ (we
have omitted them for simplicity). Moreover, $H$ denotes a general\footnote{%
As opposed to its confluent special cases.} Heun function $%
H(a,q,\alpha,\beta,\gamma,\delta,z)$ which satisfies the equation (see \cite{Heun})
\begin{equation}
\frac {d^2w}{dz^2} + \left[\frac{\gamma}{z}+ \frac{\delta}{z-1} + \frac{%
\epsilon}{z-a} \right] \frac {dw}{dz} + \frac {\alpha \beta z -q} {%
z(z-1)(z-a)} w = 0 \ ,
\end{equation}
where $\epsilon=\alpha+\beta-\gamma-\delta+1$. Let us note that for both
Heun functions in Eq. (\ref{soly}) $\epsilon=1/2$.

Before continuing, we recall that Jacobi elliptic functions are
doubly-periodic; they have a real and an imaginary period. For Lam\'e
functions to be doubly-periodic $l$ must be an integer. However, for any
value of $l$, integer or not, there are infinitely many solutions with real
period $2K$ or $4K$, where $K$ denotes the quarter period integral which is
a function of $\kappa$, the elliptic modulus, i.e.,
\begin{equation}
K(\kappa)=\int_0 ^{\pi/2}\frac{\mathrm{d}\varpi}{\sqrt{%
1-\kappa^2\sin^2\varpi}} \ .
\end{equation}
Henceforth, we fix the integration constant $\kappa$ in terms of the length of the space-time cylinder by
\begin{equation}
r_f-r_i= 2K(\kappa) \ .
\end{equation}

\subsection{Non-constant Higgs profile: The kink case}

In the previous section, we showed that for non-constant Higgs profiles
solving the complete Georgi-Glashow model field equations leads generically
to a Lam\'e equation. However, a very special case arises when in Eq. %
\eqref{gg3.2} one considers $I_{0}=0$ or, equivalently $\kappa=1$ in Eq. %
\eqref{hexplicit}. In this case, the Higgs profile becomes a kink 
\begin{equation}  \label{kink}
h(u)=vev\tanh u \ ,
\end{equation}
as can be seen from equations Eqs. \eqref{hexplicit} and \eqref{Ku}. Notice that the
kink is asymptotically constant, i.e., when $u\to\pm\infty$ then $h(u)\to\pm
vev$. Thus, the kink case is connected to both of our previously examined
cases, constant and non-constant profiles.

\subsubsection{Mapping with the P\"oschl-Teller equation}

Evaluating Eq. \eqref{lame} at $\kappa=1$ yields 
\begin{equation}
\frac{\mathrm{d}^2y}{\mathrm{d} u^2}-l(l+1)\tanh^2 (u) \,y=0 \ ,
\end{equation}
which can be written as 
\begin{equation}
-\frac{1}{2}\frac{\mathrm{d}^2y}{\mathrm{d} u^2}-\frac{l(l+1)}{2}\mathrm{sech%
}^2 (u) \,y= -\frac{l(l+1)}{2}y\ .  \label{PT}
\end{equation}
This is a one-dimensional Schr\"odinger equation with a P\"oschl-Teller
potential (see \cite{barut}). The solutions of which are known to be Legendre functions of the
form $P^{\nu}_{l}(\tanh(u))$. However, just as Eq. \eqref{lame} is not the most
general Lam\'e equation also Eq. \eqref{PT} is not the most general
P\"oschl-Teller equation. In this case, we are restricted by 
\begin{equation}
\nu^2=l(l+1) \ .
\end{equation}
Notice that as $u\to\pm\infty$ then $\tanh(u)\to\pm1$. This is problematic
for us as Legendre functions are generically singular at the points $(-1,1,\infty)$. In Quantum Mechanics this issue is resolved by quantization
conditions on $l$ and $\nu$. These conditions guarantee that the Legendre
functions vanish at the boundary, as they are interpreted as wave functions.
However, under our current restriction we can only employ one quantization
condition. As a consequence, solutions $y(u)$ can be regular only at plus or
minus infinity, but not both. 

From its definition in Eq. \eqref{newchange}, we see that $\Gamma(\alpha)
$ is bounded from above, which also applies for $y(u)$. This is incompatible
with the Legendre functions in the general solution of Eq. \eqref{PT}. To
resolve this issue, we consider the space-time cylinder as semi-infinite,
meaning 
\begin{equation}
0<r<\infty \ ,
\end{equation}
where we fix the origin at the bound of $y(u)$. The solutions 
\begin{equation}
y(u)= P^{\sqrt{n(n+1)}}_n(-\tanh u) \ ,
\end{equation}
all fulfill our desiderata whenever $n$ is an integer.     

It is interesting to note that when the Chern-Simons term is included (see Section 6) it is natural to expect that only semi-infinite cylinders are allowed since the Chern-Simons coupling introduces exponential terms which decay in one direction but not in the other. What is slightly surprising is that a similar behavior is also present without the Chern-Simons term. In a sense, in the limit in which the cylinder has an infinite volume, the theory feels the presence of the Chern-Simons term even if it is not included directly in the action.

\subsection{Non-Abelian magnetic flux}

Now we discuss the resulting non-Abelian flux. First, being in a cylinder, it is
clear that we should require periodic boundary conditions in the $\phi $
direction (with period $2\pi $) for the non-Abelian field strength and for
the energy-momentum tensor. As we mentioned before, the ansatz in Eq. (\ref{opt2.1}%
) automatically satisfies this condition for $p$ an integer number. 
One can check directly that the flux in Eq. \eqref{magneticflux} for the configurations defined by the
ansatz in Eqs. \eqref{generic1}, \eqref{generic2}, \eqref{generic3}, \eqref{generic5} and \eqref{opt2.1} is given
by 
\begin{gather}
\Psi ^{1}_{\text{M}}=\Psi ^{2}_{\text{M}}=0\ ,\quad \quad \Psi ^{3}_{\text{M}}=\int dr \chi(r) \ ,  \label{fs2}
\end{gather}%
where 
\begin{equation} \label{chi}
 \chi(r) = -2kp\pi \frac{\sin
(\alpha )(\sin ^{2}(\alpha )+\cos (\alpha )\sqrt{k+\cos ^{2}(\alpha )})}{%
(k+\cos ^{2}(\alpha ))^{\frac{3}{2}}}\alpha ^{\prime } \ ,
\end{equation}
and we have integrated in the coordinate $\phi $ and used the
relation $\lambda =\lambda (\alpha )$ in Eq. \eqref{changeofvariable1}. 
\begin{figure}[ht]
\centering
\includegraphics[scale=0.8]{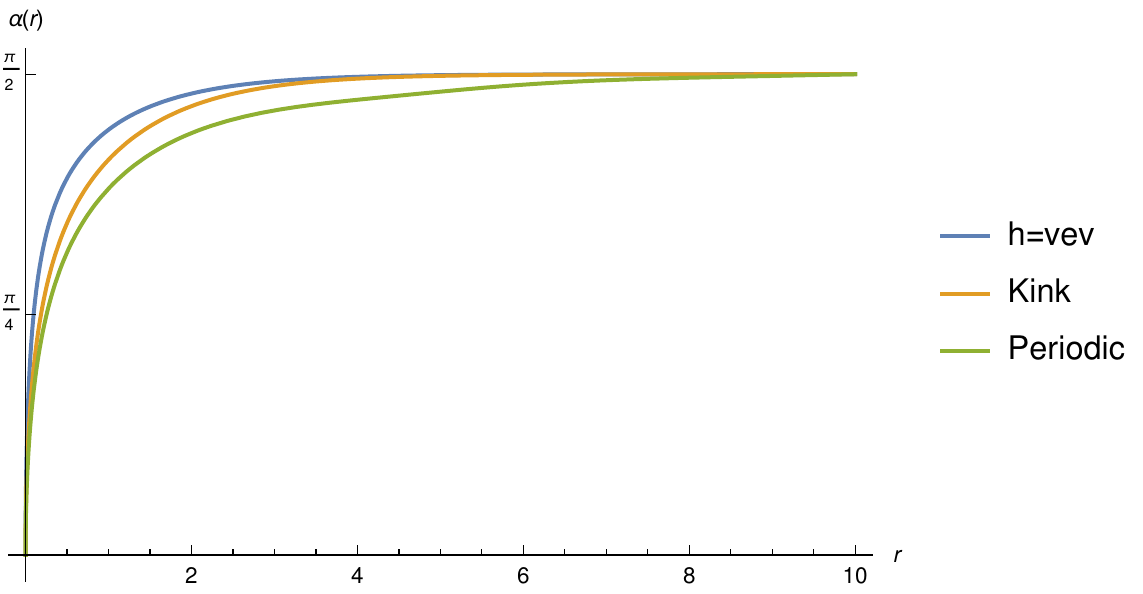} 
\caption{The regular $\alpha$ profiles for the solutions with constant Higgs, the kink case and the periodic case for the boundary conditions in Eq. \eqref{bc3}. Here we have considered $e=\gamma=vev=1$,
$R=L=p=1$, $k=0.2736$ and $\kappa=1/2$ for the periodic solution.}
\label{profiles}
\end{figure}
It is important to note that the non-Abelian flux is the same for all the configurations constructed here, that is with and without a Higgs field. Although
in the Georgi-Glashow model the Higgs field $h$ is implicit in the
solutions for the profile $\alpha $ (see for instance Eq. \eqref{ggvev3}), the final expression
for the non-Abelian flux does not depend on $h$ since the integration can
be carried out in $\alpha$ instead of the $r$ coordinate. This is due to
the fact that the integrand that appears in the non-Abelian
flux in Eq. \eqref{chi} has a global factor $\alpha ^{\prime }$. 

Now, considering the boundary conditions in Eq. \eqref{bc3}, the non-null non-Abelian magnetic flux turns out to be 
\begin{equation*}
\Psi ^{3}_{\text{M}}=\frac{p\pi }{\sqrt{k+1}}\biggl(2+2k+k\sqrt{k+1}\log (k+1)-2k\sqrt{%
k+1}\log (\sqrt{k+1}+1)\biggl)\ .
\end{equation*}%
Imposing the condition of a quantized magnetic flux, that is 
\begin{equation}
 \Psi ^{3}_{\text{M}} = np\ , \qquad \text{with} \quad n\in \mathbb{Z} \ ,  
 \label{psi=np}
\end{equation}
we get to the following (first) allowed values of the $k$
constant: 
\begin{gather*}
k_{1}=19.4003\ ,\quad k_{2}=2.62628\ ,\quad k_{3}=0.273648\ ,\quad
k_{4}=-0.42079\ ,\quad k_{5}=-0.698148\ , \\
k_{6}=-0.828956\ ,\quad k_{7}=-0.897471\ ,\quad k_{8}=-0.936029\ ,\quad
k_{9}=-0.958882\ .
\end{gather*}
Notice that in our convention (see Eqs. \eqref{I} through \eqref{EqH}) the Yang-Mills coupling constant $e$, appears in the action and not in front of commutators. In the alternative convention, magnetic flux is not dimensionless and the equivalent of Eq. \eqref{psi=np} is a quantization in terms of $1/e$.

\begin{figure}[ht]
\centering
\includegraphics[scale=0.7]{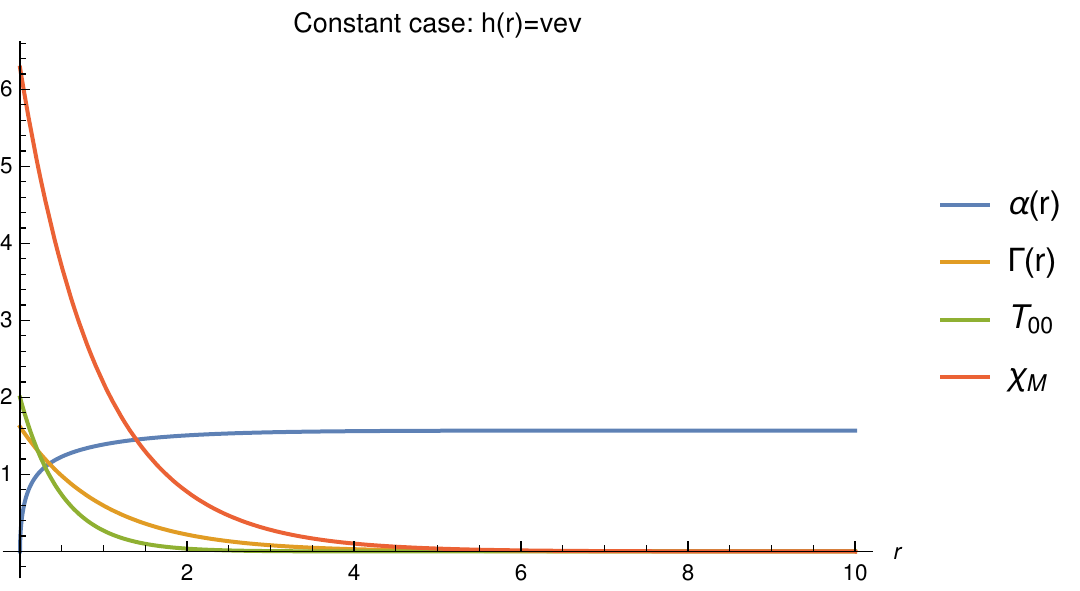} \ \qquad %
\includegraphics[scale=0.7]{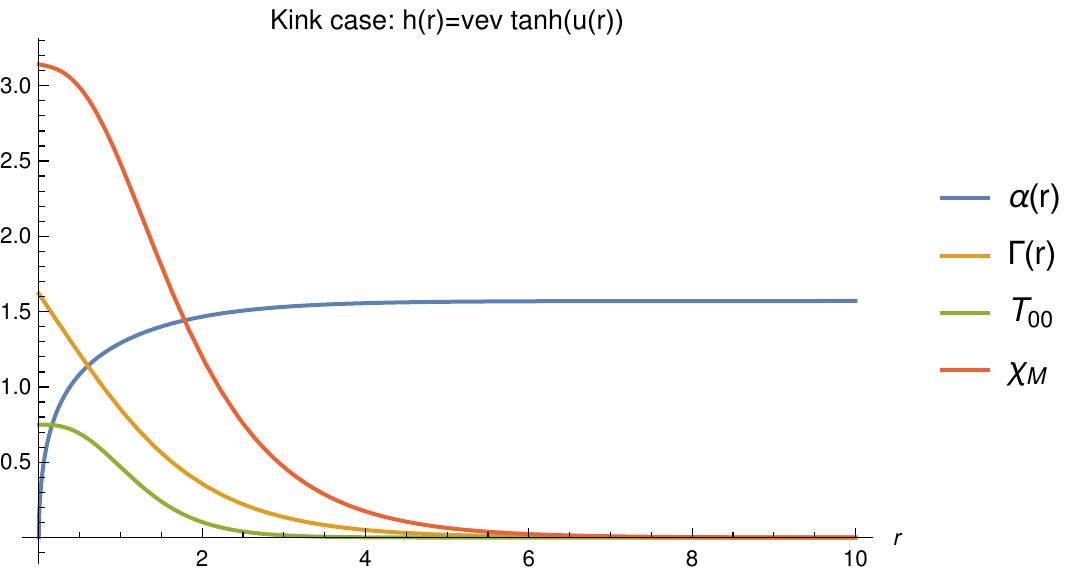} \ \qquad
\includegraphics[scale=0.7]{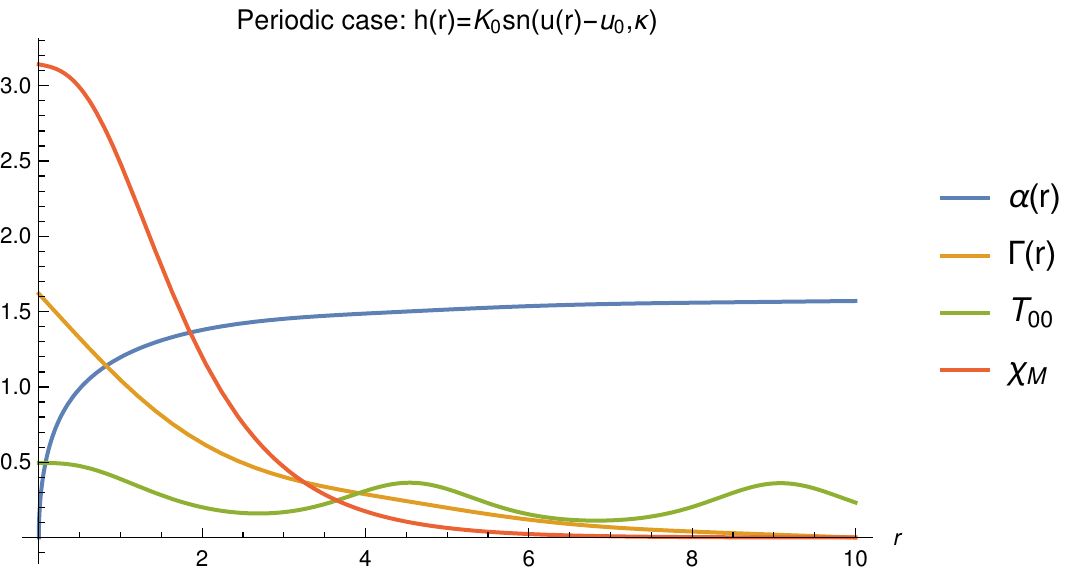}
\caption{Behavior of $\alpha(r)$ and $\Gamma(r)$ profiles as well as the energy density and the non-Abelian density flux 
for the solutions with constant Higgs, the kink case and the periodic case. Here we have considered $e=\gamma=vev=1$,
$R=L=p=1$, $k=0.2736$ and $\kappa=1/2$ for the periodic solution.}
\label{todas}
\end{figure}

\begin{figure}[ht]
\centering
\includegraphics[scale=0.9]{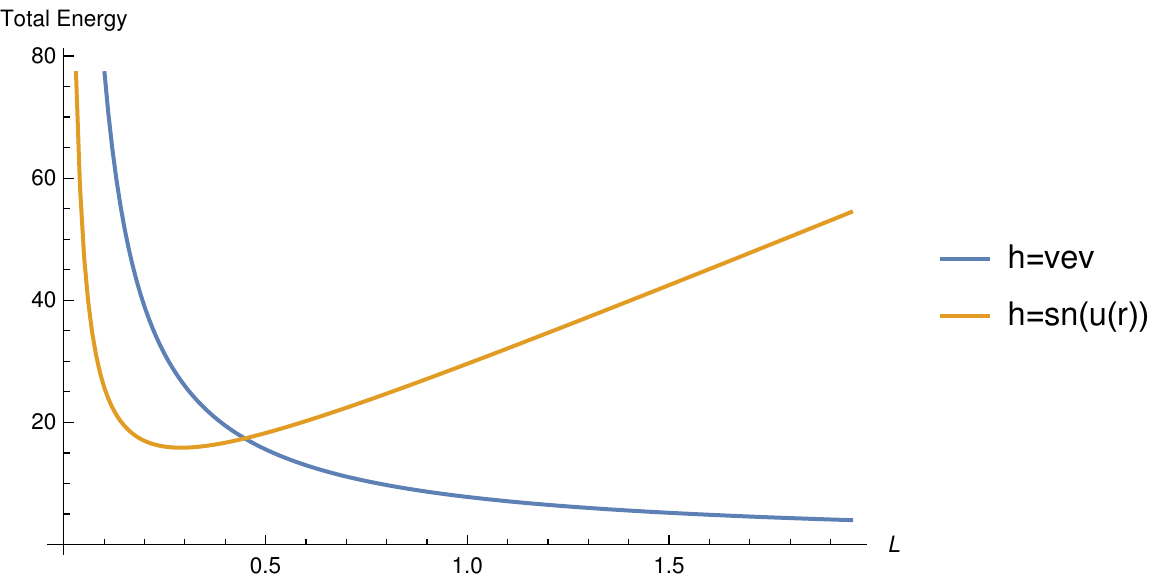} 
\caption{Energy comparison between the solutions with finite height. The length of the tube in which the condensate
is confined determines that for large values of the volume of the cylinder the energetically favored configuration is the one with a constant Higgs profile, while,
for small volumes is the one with the Higgs profile given by $h(r)=K_0 \sn(u,\kappa)$.}
\label{Comparison}
\end{figure}

With the set of values for $k$ displayed above, we can now plot some relevant physical quantities. In Fig. \ref{profiles} we show the $\alpha$ profile for all the cases when the Higgs
field is present. We can see that in all these cases, we are able to obtain regular solutions for the boundary conditions in Eq. \eqref{bc3} imposed by the condition of a non-vanishing topological charge. In Fig. \ref{todas} we show the profile, the $\Gamma$ function, the energy density and the $\chi$ function in the non-Abelian flux for the three cases in the Georgi-Glashow model. We see that in all the cases the energy as well as the non-Abelian flux are concentrated at the origin of the cylinder. 
In Fig. \ref{Comparison} we show that finite density transitions exist between the two configurations of finite height, namely the constant Higgs case and the periodic case, and this transition depends on the length of the tube in which the condensate is confined. For large values of the volume of the cylinder the energetically favored configuration is the one with a constant Higgs profile, while,
for small volumes is the periodic case. Note that it makes sense to compare the energy of these configurations since both configurations have the same magnetic flux, as can be seen from Eqs. \eqref{fs2} and \eqref{chi}.  

\section{Small fluctuations and stability}


In this section, we analyze the linearized Yang-Mills-Higgs equation on the
non-homogeneous condensate constructed in the previous sections. Since the
full stability analysis of the non-homogeneous condensates 
is extremely complicated even from the numerical viewpoint (as one should
analyze a coupled system of twelve partial differential equations in a
non-trivial background solution) we consider here the simplest
non-trivial perturbations which are under some analytic control. It is worth emphasizing that the linear stability analysis in the present section only concerns the behavior of the analytic solutions constructed in the previous sections under small perturbations. Consequently, in the present section the term "stable" actually means linearly stable. The situation should be contrasted with BPS solutions which are stable also at non-linear level. This family
of perturbations of the Yang-Mills and Higgs profiles are 
\begin{align}
h(r)\ \longrightarrow \ & h(r)+\epsilon P_{1}(r)\ , \\
\Gamma (r)\ \longrightarrow \ & \Gamma (r)+\epsilon P_{2}(r)\ ,
\end{align}%
with $\epsilon \ll 1$. The fluctuation's operator $\widehat{O}$ corresponding
to the perturbations defined above is:%
\begin{equation}
\widehat{O}\left( 
\begin{array}{c}
P_{1} \\ 
P_{2}%
\end{array}%
\right) =\left( 
\begin{array}{c}
-\frac{d^{2}P_{1}}{dr^{2}}+\gamma R^{2}(3h^{2}-vev^{2})P_{1} \\ 
-\frac{d^{2}P_{2}}{dr^{2}}+e^{2}R^{2}h^{2}P_{2}+2e^{2}R^{2}h\Gamma P_{1}%
\end{array}%
\right) \ .  \label{defO1}
\end{equation}
Therefore, a necessary condition for stability is that the operator $%
\widehat{O}$ should only possess non-negative eigenvalues. Thus, we
analyze the linear equations here below 
\begin{align}
-\frac{d^{2}P_{1}}{dr^{2}}+\gamma R^{2}(3h^{2}-vev^{2})P_{1}& =EP_{1}\ ,
\label{P1} \\
-\frac{d^{2}P_{2}}{dr^{2}}+e^{2}R^{2}h^{2}P_{2}+2e^{2}R^{2}h\Gamma P_{1}&
=EP_{2}\ ,  \label{P2}
\end{align}%
and we discuss in which of the cases discussed above $E$ is
non-negative. The proper boundary conditions in the $r$ coordinate for the
perturbations $P_{1}$ and $P_{2}$ must be chosen in such a way that the
topological charge and the non-Abelian magnetic flux must not change at order $%
\epsilon $ (otherwise $P_{1}$ and $P_{2}$ would not be small perturbations).
Eq. \eqref{P2} shows that it is convenient to solve Eq. \eqref{P1} first.

\subsection{Radial perturbations of the condensates with constant Higgs
profile}

Let us begin by considering a constant Higgs profile $h=\pm vev$ in a tube of
finite size. In this case, Eq. \eqref{P1} becomes 
\begin{equation}
P_{1}^{\prime \prime }+\omega _{1}^{2}P_{1}=0 \ ,
\end{equation}%
where 
\begin{equation}
\omega _{1}^{2}=E-2\gamma R^{2}vev^{2}.
\label{w1}
\end{equation}%
Indeed, if $\omega _{1}^{2}$ is not positive, then the perturbation $P_{1}$
could not satisfy periodic boundary conditions in $r$ (on the other hand,
when $P_{1}$ satisfies periodic boundary conditions, the topological charge
and the non-Abelian magnetic flux do not change at order $\epsilon $).
Consequently, we must have $E>2\gamma R^{2}vev^{2}$, so that the necessary
condition for stability discussed above is satisfied.

Now that we have established that radial perturbations of the profile are
harmonic, we move on to Eq. \eqref{P2} which requires we know both $%
P_{1}(r)$ and $\Gamma (r)$. By proceeding as above we find that 
\begin{equation}
P_{2}^{\prime \prime }+\omega _{2}^{2}P_{2}=F(r) \ ,
\label{eqp2}
\end{equation}%
which is the equation of an undamped driven harmonic oscillator whose
angular frequency is 
\begin{equation}
\omega _{2}^{2}=E-e^{2}R^{2}vev^{2} \ .
\end{equation}%
The driving force is partly described by $P_{1}$, which is sinusoidal, and
by $\Gamma $ which for constant $h=\pm vev$ is exponential; see Eq. %
\eqref{almostfinal}. Concretely, the driving force is 
\begin{equation}
F(r)=e^{\pm eR(vev)r}(A\cos (\omega _{1}r)+B\sin (\omega _{1}r)) \ ,
\end{equation}%
where $A$ and $B$ are fixed by the boundary conditions for $\Gamma $
and $P_{1}$.

The general solution of Eq. \eqref{eqp2} is
\begin{equation}
 P_2=C_1\cos (\omega_{2}r) + C_2\cos (\omega_{2}r)+\frac{ e^{ eR h r}[(A\gamma hR -Be\omega_{1})\cos (\omega_{1} r)+(B\gamma hR +Ae\omega_{1})\sin (\omega_{1} r)]}{2hR(e^2\omega_{1}^{2}+\gamma^2R^2h^2)} \ ,
\end{equation}
where $h=\pm vev$ in this case. Notice that perturbations never explode as 
the denominator of the particular solution is always positive.
By choosing boundary conditions for $P_2$ such that vanishes at both ends of the tubes, we fix the integration constants $C_1$ and $C_2$ in terms of $A$ and $B$. 
This choice guarantees that 
the charges do not change with the perturbation.

\subsection{Radial perturbations of the condensates with non-constant Higgs
profile}

When the Higgs profile is not constant then radial perturbations
become quite unmanageable. However, for the semi-infinite kink case we mention
the following. Carrying out the change of variable in Eq. (\ref{Ku}) allows for Eq. (\ref{P1}) to be written as
\begin{equation}
-\frac{\mathrm{d}^{2}p_{1}}{\mathrm{d}u^{2}}+6\kappa ^{2}\mathrm{sn}%
^{2}(u,\kappa )\,p_{1}=\tilde{E}p_{1}\ ,  \label{Lame2}
\end{equation}%
where $P_{1}(r(u))=p_{1}(u)$ and 
\begin{equation}
\tilde{E}=\frac{\kappa^2 +1}{\gamma R^{2}(vev)^{2}}\left[ E+\gamma
R^{2}(vev)^{2}\right] \ .
\end{equation}
One can see that the perturbation is governed by a Lam\'{e} equation with $l=2$, and whose general solution is given in terms of Heun functions. 
For the semi-infinite kink case (that is, setting $\kappa=1$), Eq. (\ref{Lame2}) becomes a P\"{o}schl-Teller equation: 
\begin{equation}
-\frac{\mathrm{d}^{2}p_{1}}{\mathrm{d}u^{2}}-6\mathrm{sech}%
^{2}(u)\,p_{1}=\left[ -4+\frac{2E}{\gamma R^{2}(vev)^{2}}\right] p_{1}\ .
\end{equation}%
In this case, using known results on the P\"{o}schl-Teller equation \cite{barut} one can show that $E\geq 0$, so that also in this
case the non-homogeneous condensate
is stable under the perturbation defined here above.


\section{Analytic non-homogeneous condensates in the
Yang-Mills-Higgs-Chern-Simons theory}


In previous sections, we have seen that Eq. \eqref{changeofvariable1} is
applicable for both constant and non-constant Higgs profiles. Remarkably,
when the Yang-Mills-Higgs action is additionally coupled to Chern-Simons
theory the equations of motion are still compatible with our general
approach, as shown below.

Also in this case, for the ansatz defined in Eqs. \eqref{generic1}, \eqref{generic2},
\eqref{generic3}, \eqref{generic5}, \eqref{generic4} and \eqref{opt2.1} together with the relation between $\lambda $
and $\alpha $ in Eq. \eqref{changeofvariable1}, the complete set of
Yang-Mills-Higgs-Chern-Simons field equations in Eqs. \eqref{Eq2} and %
\eqref{EqH} are reduced to just one equation for the soliton profile, that
is 
\begin{equation}
\alpha ^{\prime \prime }+\left[ \cot \alpha +\frac{3\sin \alpha \cos \alpha 
}{\cos ^{2}\alpha +k}\right] \alpha ^{\prime 2}+\frac{e^{2}R^{2}h^{2}}{k}%
\cot \alpha (\cos ^{2}\alpha +k)-\mu R\alpha ^{\prime }=0\ ,  \label{cs}
\end{equation}%
while, the Higgs potential $h(r)$ has the general solution in Eq. %
\eqref{hexplicit}.

As we did before, this equation can be simplified by using the change in
Eqs. \eqref{newchange} and \eqref{newchange2} leading to the generalization of %
Eq. \eqref{almostfinal}: 
\begin{equation}
\Gamma ^{\prime \prime }-(e^{2}R^{2}h^{2})\Gamma -\mu R\Gamma ^{\prime }=0\ .
\end{equation}%
As the Higgs profile $h$
acquires its simplest form in terms of the variable $u$, cf. Eq. \eqref{Ku}, we
make this change and arrive at 
\begin{equation}
\frac{\mathrm{d}^{2}y}{\mathrm{d}u^{2}}-l(l+1)\kappa ^{2}\mathrm{sn}%
^{2}(u)y-m\sqrt{1+\kappa^2 }\frac{\mathrm{d}y}{\mathrm{d}u}=0 \ ,
\end{equation}%
where $m=\frac{\mu}{vev\sqrt{\gamma }}$ and $l(l+1)=\frac{2e^{2}}{\gamma}$, as defined
earlier. By reparameterizing $y$ as 
\begin{equation}
y(u)=e^{\frac{m}{2}\sqrt{1+\kappa^2 }u}Y(u) \ ,
\end{equation}%
we see that $Y(u)$ is a Lam\'{e} function, as it satisfies 
\begin{equation}
\frac{\mathrm{d}^{2}Y}{\mathrm{d}u^{2}}-l(l+1)\kappa ^{2}\mathrm{sn}%
^{2}(u)Y-\frac{1}{4}m^{2}(1+\kappa^2 )Y=0 \ .
\end{equation}%
Moreover, when $\kappa =1$ we have a P\"{o}schl-Teller equation 
\begin{equation}
-\frac{1}{2}\frac{\mathrm{d}^{2}Y}{\mathrm{d}u^{2}}-\frac{l(l+1)}{2}\mathrm{%
sech}^{2}(u)Y=-\frac{1}{2}\left[ l(l+1)+\frac{1}{2}m^{2}\right] Y \ .
\end{equation}%
We also remark that, similar to the case $m=0$, of previous sections,
desired solutions on a semi-infinite space-time cylinder are obtained by
integer values of $l$. Interestingly enough, $m$ can take arbitrary values.
In other words, 
\begin{equation}
y(u)=e^{\frac{m}{\sqrt{2}} u} P_{n}^{\sqrt{n(n+1)+m^{2}/2}}(-\tanh u) \ ,
\end{equation}%
where $m$ is real but $n$ is an integer and $P$ is a Legendre function. To the best of the authors' knowledge,
this is the first family of analytic topologically non-trivial solutions of
non-homogeneous condensates in the Yang-Mills-Higgs-Chern-Simons theory. We
analyze the physical properties of these solutions in a forthcoming
paper. Here, we only remark that, due to the presence of the
Chern-Simons term, the factor $e^{\frac{m}{\sqrt{2}}u}$ (which appears in the
solution here above) is well-defined either on finite intervals or (if one
is interested in the infinite volume limit) when $-\infty < u\leq 0$.
Hence, in the case in which the Chern-Simons coupling is included, one can
only have semi-infinite tubes.

\section{Conclusions}

Using a non-spherical generalization of the usual hedgehog ansatz, the first
analytic examples of non-homogeneous condensates, both in the
$(2+1)$-dimensional Georgi-Glashow model as well as in the
Yang-Mills-Higgs-Chern-Simons theory at finite density have been
constructed. These exact configurations live within a cylinder which can
have either finite height or can be infinitely long on one side, being the maximum of the energy density located at the origin of the tube.  
These non-homogeneous condensates possess a (novel) non-trivial topological charge, in such a way that the condensates
cannot decay into the trivial vacuum. Such charge does not coincide with the non-Abelian magnetic flux, which usually plays the role of
the topological charges in gauge theories. 
Requiring the quantization of the non-Abelian flux one of the free parameters characterizing 
the ansatz for the non-Abelian gauge field can be fixed. We show that depending on the length of the cylinder,
finite density transitions occur. In particular, for large values of $L$ the energetically favored
configuration is the one with a constant Higgs profile, while, for small values is the one with the Higgs profile given by an elliptic function. 
Also, we have derived some necessary conditions in order to have stable condensates under radial perturbations.
These surprising results open the possibility to study the dynamics of
Quarks moving in these topologically non-trivial condensates with analytic
tools. We hope to come back on this important issue in a future publication.

\subsection*{Acknowledgments}

F. C. has been funded by Fondecyt Grants 1200022. M. L. is funded
by FONDECYT post-doctoral Grant 3190873. A.V. is funded by FONDECYT post-doctoral Grant 3200884.
D.F. is supported by a CONACYT postdoctoral fellowship. This work has been
partially funded by CONACYT Grant No. A1-S-11548. The Centro de Estudios
Cient\'{\i}ficos (CECs) is funded by the Chilean Government through the
Centers of Excellence Base Financing Program of Conicyt.

\section{Appendix: Some useful tensors}

In this Appendix, we explicitly show some quantities that help to clarify
the construction of our analytic solutions starting from the ansatz in Eqs. 
\eqref{metric1}, \eqref{ranges}, \eqref{generic1}, \eqref{generic2}, \eqref{generic3}, \eqref{generic5}, \eqref{generic4} and \eqref{opt2.1}.

First of all, the matrices $U$ and $\varphi$ are explicitly given by 
\begin{equation*}
U= 
\begin{pmatrix}
\cos(\alpha) & ie^{-i \Phi}\sin(\alpha) \\ 
i e^{i\Phi} \sin(\alpha) & \cos(\alpha)%
\end{pmatrix}
\ , \qquad \varphi= 
\begin{pmatrix}
0 & ie^{-i \Phi}h(r) \\ 
i e^{i\Phi} h(r) & 0%
\end{pmatrix}
\ .
\end{equation*}%
For the computation of relevant physical quantities, it is convenient to define the following functions 
\begin{align*}
f_1(r) \ = \ & -2\cos(\alpha) \alpha^{\prime }(\lambda-1)\lambda
+\sin(\alpha)\lambda^{\prime }\ , \\
f_2(r) \ = \ & 2\sin(\alpha) \alpha^{\prime }(\lambda-1)\lambda
+\cos(\alpha)\lambda^{\prime }\ ,
\end{align*}
and also 
\begin{align*}
g_1(r) \ = \ & 2\sin^2(\alpha) \lambda -1 \ , \\
g_2(r) \ = \ & \sin(\alpha) \lambda \ .
\end{align*}
From the above, one can check that the non-vanishing components of
the non-Abelian field strength are 
\begin{equation*}
F_{tr}=p 
\begin{pmatrix}
-i f_1(r) & -e^{-i\Phi}f_2(r) \\ 
e^{i\Phi} f_2(r) & i f_1(r)%
\end{pmatrix}%
\sin(\alpha) \ , \qquad F_{r\phi} = L F_{tr} \ .
\end{equation*}%
The non-Abelian character of our solutions is made manifest by non-null
commutators that appear in the field equations in Eqs. \eqref{Eq2} and %
\eqref{EqH}. Indeed, by defining the following tensors 
\begin{equation*}
P^\mu = [A_\nu,F^{\mu\nu}] \ ,
\end{equation*}
\begin{equation*}
J^\mu = [\varphi,D^\mu \varphi] \ ,
\end{equation*}
a direct calculation shows that 
\begin{equation*}
P^{t}=\frac{2p}{L} 
\begin{pmatrix}
-i f_2(r) & e^{-i\Phi}f_1(r) \\ 
-e^{i\Phi} f_1(r) & i f_2(r)%
\end{pmatrix}%
\sin(\alpha)\alpha^{\prime} \ , \qquad P^{\phi}= \frac{1}{L}P^{t} \ ,
\end{equation*}%
and 
\begin{equation*}
J^{t}=\frac{2p}{L} 
\begin{pmatrix}
i g_1(r) & e^{-i\Phi}g_2(r) \\ 
-e^{i\Phi} g_2(r) & i g_1(r)%
\end{pmatrix}%
h^2 \ , \qquad J^{\phi} = \frac{1}{L} J^{t} \ .
\end{equation*}%
Note that the tensors $P^\mu$ and $J^\mu$ are zero along the radial components.

\end{document}